\documentclass[runningheads]{llncs}
\usepackage[T1]{fontenc}
\usepackage{graphicx}
\usepackage{algorithm}
\usepackage{algpseudocode}
\newcommand\given[1][]{\:#1\vert\:}

\usepackage{cite}
\usepackage{hyperref}

\graphicspath{ {figures/} }

\begin{document}
\title{Towards Automatic Abdominal MRI Organ Segmentation: Leveraging Synthesized Data Generated From CT Labels}

%%% Anonymized for MICCAI %%% 
% \author{First Author\inst{1}\orcidID{0000-0000-0000-0000} \and
% Second Author\inst{1}\orcidID{0000-0000-0000-0000} \and
% Third Author\inst{2}\orcidID{0000-0000-0000-0000} \and 
% Fourth Author\inst{3}\orcidID{0000-0000-0000-0000} \and
% Fifth Author\inst{1}\orcidID{0000-0000-0000-0000} \and
% Sixth Author\inst{1}\orcidID{0000-0000-0000-0000}
% \institute{Anonymous organization \and
% Anonmymous organization \and
% Anonymous organization}

\author{Cosmin Ciausu\inst{1} \and
Deepa Krishnaswamy\inst{1} \and
Benjamin Billot\inst{2} \and 
Steve Pieper\inst{3} \and
Ron Kikinis\inst{1} \and
Andrey Fedorov\inst{1}
}
\authorrunning{C. Ciausu et al.}
\institute{Brigham and Women’s Hospital, Boston MA 02115, USA, \and
Massachusetts Institute of Technology, Cambridge MA 02139, USA, \and
Isomics, Cambridge MA 02138, USA}

\maketitle              % typeset the header of the contribution
%

%%%%%%%%%%%%%%%%%%%%%%%%%%%%%%%%%%%%%%%%%%%%%%%%%%%%%%%%%%%%%%%%%%%%%%%%%
\begin{abstract}
%The abstract should briefly summarize the contents of the paper in
% In order to develop and validate artificial intelligence methods, it is of utmost importance to have access to both publicly available imaging data with corresponding and annotations accompanying those images. Unfortunately, there is a lack of availability of such annotations, as they can be difficult and time-consuming to obtain., Deep learning has shown great promise in the ability to automatically annotate organs in magnetic resonance imaging (MRI) scans. However, despite advancements in the field, the ability to accurately segment abdominal organs across a domain gap remains largely unsolved. In part, this may be explained by the much greater variability in image appearance and severely limited availability of training labels in the publicly available datasets. We leverage a domain randomization strategy for generating synthetic data from CT label maps, with the ability to segment both MRI and CT volumes. For training data we make use of automated label maps. The synthetic data is created from the label maps by randomizing the intensities, applying spatial transformations and artefacts, and by varying the resolution. Synthetic data is generated on the fly and used to train a U-Net model. To evaluate the approach, we use publicly available data. The method shows promise in using label maps from CT data to train a model that can be used for segmentation of abdominal organs in MR.

Deep learning has shown great promise in the ability to automatically annotate organs in magnetic resonance imaging (MRI) scans, for example, of the brain. However, despite advancements in the field, the ability to accurately segment abdominal organs remains difficult across MR. In part, this may be explained by the much greater variability in image appearance and severely limited availability of training labels. The inherent nature of computed tomography (CT) scans makes it easier to annotate, resulting in a larger availability of expert annotations for the latter. We leverage a modality-agnostic domain randomization approach, utilizing CT label maps to generate synthetic images on-the-fly during training, further used to train a U-Net segmentation network for abdominal organs segmentation. Our approach shows comparable results compared to fully-supervised segmentation methods trained on MR data. Our method results in Dice scores of 0.90 ± 0.08 and 0.91 ± 0.08 for the right and left kidney respectively, compared to a pretrained nnU-Net model yielding 0.87 ± 0.20 and 0.91 ± 0.03. We will make our code publicly available.

\keywords{image segmentation \and domain randomization \and computed tomography \and magnetic resonance imaging \and abdominal}

\end{abstract}

%%%%%%%%%%%%%%%%%%%%%%%%%%%%%%%%%%%%%%%%%%%%%%%%%%%%%%%%%%%%%%%%%%%%%%%%%

\section{Introduction}

Accurate segmentation of abdominal organs in magnetic resonance (MR) images would be beneficial for many clinical tasks including liver volumetry \cite{Gotra2017}, kidney disease monitoring \cite{Kline2018}, adaptive radiotherapy \cite{Tetar2019}. However, manual delineations of organs by experts are often time-consuming and tedious to perform \cite{Heerkens2017}. The use of supervised convolutional neural network (CNN) methods for segmentation help alleviate the aforementioned issues, being time-efficient and robust to in-domain training data. 
\\
\indent Popular supervised CNN methods for segmentation are the U-Net architecture \cite{Ronneberger2015} and its variants, including V-Net \cite{Milletari2016}, and others built on top such as nnU-Net \cite{Isensee2021}. Specifically for abdominal segmentation, multiple approaches have been developed based on these architectures. A multi-2D slice input approach was used to train a U-Net-based neural network, segmenting ten abdominal organs \cite{Chen2020}, and others used a 3D U-Net based approach \cite{Bobo2018} for MR pancreas segmentation. Going beyond the traditional encoder-decoder segmentation models, others have integrated the use of conditional generative adversarial networks (GAN) \cite{Conze2021}. There has also been multiple challenges related to the development of supervised algorithms for multi-modality abdominal segmentation, such as AMOS \cite{Ji2022} and CHAOS \cite{Kavur2019, Kavur2020, Kavur2021}. Most of the supervised methods can adapt well to the domain of the training dataset, but can fail on out-of-distribution data \cite{Donahue2014}. This is a significant concern as MR imaging data is highly heterogeneous with regards to resolution, orientation, and soft tissue contrast. Additionally, these methods require the need for large training datasets, which may not be readily available. 
\\
\indent To rectify the need of large training datasets, techniques like data augmentation have been used to increase the heterogeneity of the data. Many techniques have been developed, ranging from basic transforms, to deformable and other learning-based methods \cite{Chlap2021, Yun2019}. Though they may work well on a downstream segmentation task within a single modality, they may suffer when used in a cross-modality setting. This has led to the development of data augmentation techniques specifically for cross-modality use cases \cite{Chen2021}. From that point, the field of domain adaptation has expanded rapidly, with the development of methods to account for domain shift between training and testing datasets \cite{Guan2021}. For instance, a cross-modality domain adaptation method such as SIFA \cite{Chen2019}, modifies the input image domain to appear like the target domain using a GAN approach. Our proposed method uses a domain randomization approach, alleviating the need for defined source and target domain. 
\\
\indent The proposed method implements a deep learning-based algorithm specifically for modality agnostic abdominal organ segmentation. The method contributes the following: 1) Leveraging publicly available computed tomography (CT) label maps for synthesizing training data for primarily MR segmentation, using a domain randomization approach 2) Extensive validation and testing of our approach on publicly available datasets 3) Analyzing the effect of the number of labels for contextual information and the level of granularity of the labels on the performance of the network 4) Comparison to multiple methods publicly available in the literature.  

%%%%%%%%%%%%%%%%%%%%%%%%%%%%%%%%%%%%%%%%%%%%%%%%%%%%%%%%%%%%%%%%%%%%%%%%%

\begin{figure}
\includegraphics[width=0.9\textwidth, trim={0cm 5cm 0cm 5.5cm},clip]{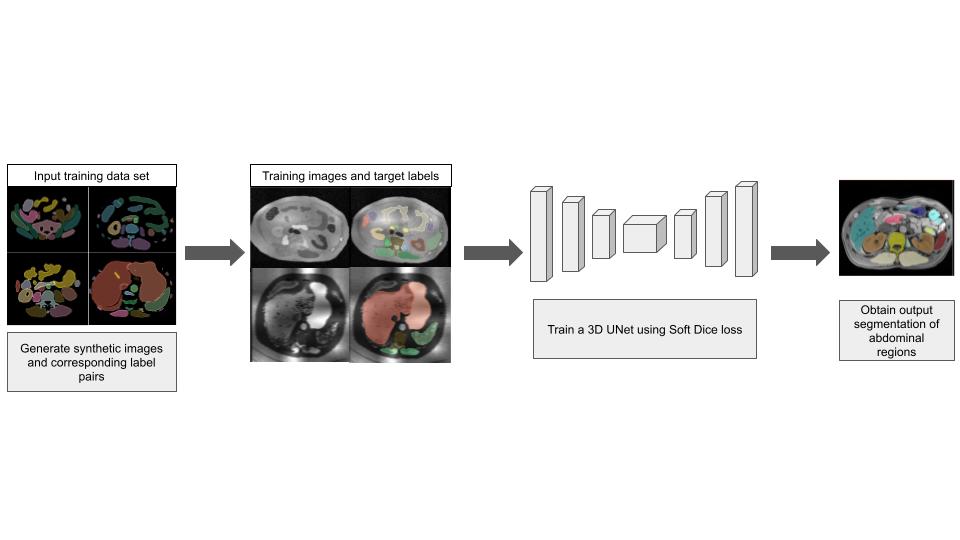}
\centering
\caption{Flowchart of the method for abdominal segmentation. The model requires as an input training label maps as input which are used to create synthetically generated images and corresponding label maps on the fly. These are used to train a U-Net model using a soft Dice loss. The final segmentation map of the anatomical structures of interests are obtained using the trained model.} \label{Figure1}
\end{figure}

% \begin{table}
% \caption{Dataset split for validation and testing, for AMOS and CHAOS. AMOS contains MR and CT modalities, and CHAOS contains T1 in phase, T2 out phase and T2 along with CT [add references later].}\label{Table1}
% \centering
% \begin{tabular}{|c|c|c|c|}
% \hline
% Dataset & Modality  & Validation subjects used & Testing subjects used \\
% \hline
% AMOS & MR & 40 & 20 \\
% AMOS & CT & 25   & 12   \\
% CHAOS & MR & 24  & 24   \\
% CHAOS & CT & 10  & 10 \\
% \hline
% \end{tabular}
% \end{table}

\section{Methods}

\subsection{Approach}

We adopt the domain randomization strategy of SynthSeg \cite{Billot2023} for the automatic segmentation of abdominal structures in MR data. The method was originally developed for the segmentation of brain structures in MR and CT volumes, and was extended to the delineation of cardiac structures \cite{Billot2023}. Using solely label maps as an input, the method generates synthetic data which is then used to train a U-Net model \cite{Ronneberger2015} for segmentation. Synthetic scans are generated by sampling a Gaussian mixture model process conditioned on the expert annotated label maps. These parameters include intensity, bias, resolution, and orientation. During the training process, volumes are generated on the fly, where the network sees randomized data at every step, thereby becoming agnostic to resolution and contrast. Please refer to the paper \cite{Billot2023} for further details of the adopted domain randomization strategy. Figure \ref{Figure1} describes the overall methodology for our proposed abdominal organ segmentation approach. 
% For each step of the training process, a random label map is selected, and a synthetic image with corresponding target label maps is created using the generative model \cite{Billot2023}. The sets of image and label map pairs are used to train U-Net \cite{Ronneberger2015} model using a soft Dice loss.

\subsection{Datasets}
\subsubsection{Training}
We use the publicly available data (v1) provided from the TotalSegmentator method for CT structure segmentation \cite{Wasserthal2023}. All of the data were resampled to 1.5\,mm, and includes 104 anatomical regions. The authors used a semi-automatic process to generate ground-truth annotations, combining AI-produced segmentations and expert feedback. For our study, we used 10 subjects for training, as it has previously been shown that the generative model performs well despite a low number of training subjects \cite{Billot2023}. We randomly chose 10 patients where the abdomen was present. 

\begin{table}
\caption{Dataset split for validation and testing, for AMOS and CHAOS, in terms of number of subjects (subjs). AMOS contains MR and CT modalities, and CHAOS contains T1 in phase, T2 out phase and T2 along with CT.}\label{Table1}
\centering
\begin{tabular}{|c|c|c|c|c|c|}
\hline
Dataset & \# MR val subjs & \# MR test subjs & \# CT val subjs & \# CT test subjs \\
\hline
AMOS \cite{Ji2022} & 40 & 20 & 25   & 12 \\
CHAOS \cite{Kavur2019, Kavur2020, Kavur2021} & 24  & 24 & 10  & 10  \\
\hline
\end{tabular}
\end{table}

\subsubsection{Validation and testing}

We used two publicly available collections for validation and testing of our approach. The AMOS dataset \cite{Ji2022} is a collection of patients with abdominal cancer or abnormalities, where both CT and MR data were collected from two medical centers using eight different scanners. The dataset includes labels for 15 abdominal organs. A coarse segmentation was first performed using a model trained on a small sample of the data, and the output of these models were refined by junior radiologists for the remaining data, and reviewed by senior radiologists. We used a subset of this data for validation and testing, as detailed in Table~\ref{Table1}. 

\indent The second publicly available dataset we used was CHAOS \cite{Kavur2019, Kavur2020, Kavur2021}, which consists of CT and MR images from healthy patients at the Dokuz Eylul University Hospital, in Izmir, Turkey. MR sequences included T1 in phase, T1 out phase and T2 SPIR, and includes labels for the liver, spleen, right kidney and left kidney, while the CT dataset only includes labels for the liver. These images were annotated from three radiologists with a majority voting procedure. We used a subset of the provided training dataset for our validation, and the provided validation set for our testing. We split the data into validation and testing cohorts as described in Table~\ref{Table1}. 

% table 1 was here before

\subsubsection{Pre-processing for target labels selection}

The original TotalSegmentator training data contains whole-body CT images and label maps, with 104 structures segmented. In our study, we only focus on the abdominal organs, therefore we cropped and/or padded our images and label volumes to a fixed size of 300x300x250. The following organs were selected for segmentation: liver, spleen, kidneys, stomach, duodenum, pancreas, gallbladder, small bowel, colon, adrenal glands, sacrum, hip bone, gluteus maximus, gluteus medius, gluteus minimus, autochthon, iliopsoas. We combined the vertebrae present in the abdominal region into a single segment for prediction. This yielded a total of 26 predicted labels (including left and right separately). 

\subsubsection{Pre-processing for synthetic data generation}
Our U-Net segmentation network is trained on synthetic scans generated from our training label maps. In order to diversify the appearance of our synthetic images, we applied a number of additional processing steps to increase the variety of details generated. From the original 10 label maps, we removed the CT table from 50\% of the subjects using a 3DSlicer extension \cite{Fedorov2012}. To further enhance the appearance of fine structures in CT scans later used for clustering, pre-processing steps including Gaussian filter blurring (\url{https://docs.scipy.org/doc/scipy/reference/generated/scipy.ndimage.gaussian_filter.html
}) and gamma-based contrast-stretching was applied to the CT images used to generate synthetic data during training. Additional labels only used during synthetic data generation were added, to add more randomization to our training images. These labels were obtained by performing intensity clustering on the background and the foreground of our CT training scans. A Gaussian Mixture Model (GMM) with {$K_{background} \in [3,4,5]$} components was fitted on the background voxels of our training CT scans. This GMM was then optimized using the Expectation-Maximization (EM) algorithm \cite{Dempster1977}, providing $K$ clusters assignments for all of our background CT images voxels. These additional background clusters were used during our synthetic data generation step. The same approach was used to produce additional foreground label clusters, for {$K_{foreground} \in [1,2,3]$} for each available segment from the TotalSegmentator label maps. Algorithm~\ref{Algorithm1} describes the process of background and foreground clustering GMM process used to generate additional labels for synthetic data generation. Using this process, we obtained 180 training label maps from the original 10 selected label maps.

\begin{algorithm}[t]
\caption{Background and foreground labels clustering used for synthetic image generation}\label{Algorithm1}
\begin{algorithmic}
\For {$BG$ in [3,4,5]}
    \State {$P(x)= \sum_{k=1}^{BG} \mathcal{N}(x \given \mu_{k},\,\sigma^{2}_{k}) \pi_{k} $} 
    \State Optimise {$\theta_{BG}  = \{ \mu_{k},\,\sigma^{2}_{k},\ \pi_{k}, k=1,...,BG  \}$} using EM algorithm
\For {$FG$ in [1,2,3]}
\For {segment in [1,...,$N_{seg}$]}
\State {$P(x_{segment})= \sum_{k=1}^{FG} \mathcal{N}(x_{segment} \given \mu_{k},\,\sigma^{2}_{k}) \pi_{k} $} 
\State Optimise {$\theta_{FG}  = \{ \mu_{k},\,\sigma^{2}_{k},\ \pi_{k}, k=1,...,FG  \}$} using EM algorithm
\EndFor
\EndFor
\EndFor
\State {$N_{Seg}$ refers the total number of labels present in a particular CT scan and corresponding label map}
\end{algorithmic}
\end{algorithm}
\subsection{Training procedure and evaluation}

Following the published method developed by \cite{Billot2023} from here (\url{https://github.com/BBillot/SynthSeg}), we trained the U-Net segmentation network on our synthetic images using the default parameters. We trained our network for two weeks using an NVIDIA A100 GPU with 40 GB RAM \cite{Stewart2015} for 100 epochs with 5000 steps per epoch. Epoch 10 was chosen for testing. Overlap metrics and distance-based metrics were computed for testing and validation of our methods, namely Dice score \cite{Dice1945} and Hausdorff distance \cite{Huttenlocher1993}.

% We computed the Dice score , which measures a degree of overlap between two sets of voxels, with 0 being no overlap and 1 being complete overlap between two sets. We also computed the Hausorff distance, a measure of the local maximum distance, where the smaller the value is in mm, the closer the two segmentations are . 

%%%%%%%%%%%%%%%%%%%%%%%%%%%%%%%%%%%%%%%%%%%%%%%%%%%%%%%%%%%%%%%%%%%%%%%%

\begin{table}[h]
\caption{Quantitative results for inference on two collections AMOS and CHAOS for each modality (MR and CT) in terms of mean Dice score (Dice) and mean Hausdorff distance (95th percentile) (HD95) in mm. The standard deviation values are also provided.}\label{Table2}
\centering
\begin{tabular}{|c|c|c|c|c|c|}
\hline
Dataset & Segment & MR & MR & CT & CT \\
\hline
& & Dice & HD95 & Dice & HD95 \\
\hline
AMOS & Liver & 0.90 ± 0.04 & 25.13 ± 26.0 & 0.91 ± 0.05 & 13.15 ± 15.4 \\ %&  
AMOS & Spleen & 0.86 ± 0.15 & 5.34 ± 6.37 & 0.78 ± 0.22 & 19.37 ± 26.20 \\ %& 
AMOS & Right kidney & 0.90 ± 0.08 & 4.24 ± 5.50 & 0.91 ± 0.05 & 2.83 ± 1.84 \\ %&
AMOS & Left kidney & 0.91 ± 0.08 & 2.92 ± 2.04 & 0.92 ± 0.03 & 3.55 ± 2.84 \\ %&
\hline
CHAOS & Liver & 0.87 ± 0.05 & 5.77 ± 3.27 & 0.91 ± 0.10 & 21.08 ± 23.95 \\%&
CHAOS & Spleen  & 0.78 ± 0.13  & 10.46 ± 15.06 & --- & --- \\ %&
CHAOS & Right kidney & 0.80 ± 0.14 & 3.53 ± 2.29 & --- & --- \\ %&
CHAOS & Left kidney & 0.68 ± 0.31 & 7.79 ± 11.19 & --- & --- \\ %&
\hline
\end{tabular}
\end{table}

\begin{figure}
\centering
\includegraphics[width=0.85\textwidth, trim={0cm 2.5cm 0cm 4.5cm},clip]{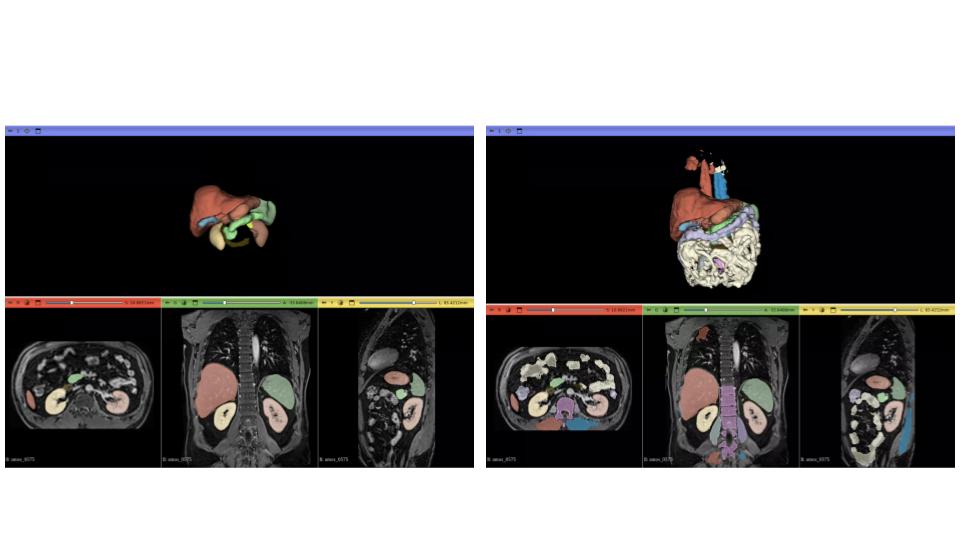}
\caption{Qualitative results of the proposed method on a subject from AMOS MR. The left view shows the expert radiologist annotations, and the right one shows the results from our baseline method. Here we have high agreement between many abdominal regions, including the liver in red, spleen in green, kidneys in yellow and brown. Our method segments 26 structures in the abdominal area.} \label{Figure2}
\end{figure}

\section{Results and Discussion}

\subsection{Experimental Results}

% \begin{figure}
% \includegraphics[width=0.9\textwidth, trim={0cm 4.2cm 1.5cm 4cm},clip]{figures/Figure2.jpg}
% \caption{Inference results for collections AMOS4 and CHAOS5–7, with their two modalities, MR and CT. The left plot displays the distributions of the Dice scores, while the right plot displays the distributions of the Hausdorff 95th percentile distances in mm. Both include the major abdominal organs, namely the liver, spleen, right kidney and left kidney.} \label{Figure2}
% \end{figure}
We performed inference on the external testing collections outlined in Table  ~\ref{Table1}. Table ~\ref{Table2} provides the quantitative results for the same collections for the liver, spleen and kidneys. We observe high Dice scores for AMOS MR and CT modalities. However, CHAOS MR yields relatively lower results than AMOS. This could be partly due to the differences between the ground truth segmentations, as both the TotalSegmentator and AMOS collections did not contain the renal cavity of the kidney, while CHAOS did. Figure ~\ref{Figure2} displays our qualitative results on one sample subject, showing the ground truth segments vs predictions from our proposed method. Please refer to the supplementary material for additional box plots results.
% Figure~\ref{Figure2} displays the Dice score results for AMOS (MR and CT modalities), as well as CHAOS (MR and CT modalities) for the liver, spleen and kidneys segments, compared to the expert annotated ground truth labels.  

\subsection{Ablation study results}

In order to assess the performance of the model and to understand the effect of the synthetic generation process, we performed a number of ablation studies. We studied the effect of foreground labels clustering and the effect of segmenting one label versus all abdominal labels. The three studies we performed were 1) No foreground clustering and segmenting all original abdominal labels, 2) Include foreground clustering and predict a single label (left and right kidneys), 3) No foreground clustering and predict a single label (left and right kidneys).
% \begin{enumerate}
%         \item No foreground clustering and segmenting all original abdominal labels 
%         \item Include foreground clustering and predict a single label (left and right kidneys)
%         \item No foreground clustering and predict a single label (left and right kidneys) 
% \end{enumerate}
\indent After examining the validation curves, epoch 5 was chosen for both A. and B. The last epoch was chosen for C. Please refer to the supplementary material for analysis of the validation loss curves. Overall our baseline method performed the best compared to the three ablation studies. Table ~\ref{Table3} shows that our baseline experiment including all abdominal organs and the additional clustering method for the synthetic generation step is significantly better than the ablation methods, across all collections. One could argue that the clustering step adds more diversity to the synthesized images used for training, and including more segmentation labels during training helps to add contextual information. Please refer to the supplementary material for qualitative results.
\begin{table}[t]
\caption{Dice score and HD95 quantitative testing results for inference on AMOS and CHAOS for MR and CT modalities. The last four rows are our baseline method. Subjects with undefined Hausdorff distance values were removed for the overall metrics calculation. Values in bold are the best performing. Kruskal-Wallis statistical testing was performed comparing our baseline approach to the three ablation studies. Values with an asterisk indicate a significant difference of p<0.01. C for the first column refers to clusters (Yes/No). The L column refers to the the number of labels, all or 1, RK refers to right kidney segmentation, LK refers to left kidney segmentation.}\label{Table3}
\centering
\begin{tabular}{|c|c|c|c|c|c|c|}
\hline
C & L & Dataset & RK Dice & RK HD95 & LK Dice & LK HD95 \\
\hline
N & 1 & AMOS MR & 0.58 ± 0.34 & 162.87 ± 42.94 & 0.25 ± 0.15 & 203.44 ± 59.35 \\ %&
N & 1 & AMOS CT &  0.0 ± 0.0 & 250.08 ± 31.09 & 0.02 ± 0.06 & 285.07 ± 19.28 \\ %& 
N & 1 & CHAOS MR & 0.29 ± 0.36 & 94.12 ± 36.03 & 0.14 ± 0.16 & 111.78 ± 15.88 \\ %&
\hline
Y & 1 & AMOS MR & 0.82 ± 0.22 & 5.46 ± 6.68 & 0.85 ± 0.19 & 5.06 ± 6.90 \\ %&
Y & 1 & AMOS CT & 0.81 ± 0.21 & 5.55 ± 8.12 & 0.84 ± 0.23 & 4.82 ± 7.36 \\ %& 
Y & 1 & CHAOS MR & 0.51 ± 0.30 & 12.48 ± 10.73 & 0.76 ± 0.27 & 5.98 ± 7.20 \\% &
\hline
N & all & AMOS MR &  0.81 ± 0.27 & 8.05 ± 9.89 & 0.87 ± 0.08 & 6.65 ± 4.79 \\% & 
N & all & AMOS CT &  0.76 ± 0.26 & 10.69 ± 16.03 & 0.68 ± 0.28 & 37.75 ± 97.07 \\% &
N & all & CHAOS MR & 0.63 ± 0.25 & 7.88 ± 6.75 & 0.62 ± 0.26 & \textbf{6.73 ± 4.90} \\% & 
\hline
\hline
Y & all & AMOS MR & \textbf{0.90 ± 0.08*} & \textbf{4.24 ± 5.50*} & \textbf{0.91 ± 0.08*} & \textbf{2.92 ± 2.04*} \\ % & 
Y & all & AMOS CT &  \textbf{0.91 ± 0.05*} & \textbf{2.83 ± 1.84*} & \textbf{0.92 ± 0.03*} & \textbf{3.55 ± 2.84*} \\ %& 
Y & all & CHAOS MR & \textbf{0.80 ± 0.14*} & \textbf{3.53 ± 2.29*} & \textbf{0.68 ± 0.31*} & 7.79 ± 11.19* \\% & 
\hline
\end{tabular}
\end{table}

% \begin{figure}
% \includegraphics[width=\textwidth, trim={0cm 2cm 0cm 1cm},clip]{figures/Figure4.jpg}
% \caption{This figure displays the qualitative results of the kidney segmentation for the baseline method along with the three ablation studies models. The ground truth expert segmentations are given by the thick segment boundary line, while the AI predictions are filled. The right kidney is green, and the left kidney is purple. Yellow arrows indicate missing segments or incorrect segmentations. Note the higher overlap of the expert segmentations versus our AI models for methods that include all labels.} \label{Figure4}
% \end{figure}

\subsection{Comparison to publicly available methods}

We also compare our baseline method to two other publicly available abdominal segmentation methods. We use data from 23 patients from the TCGA-LIHC collection \cite{Clark2013, Erickson2016} as part of NCI Imaging Data Commons (IDC) \cite{Fedorov2023}, which have AI-generated annotations of the liver from BAMF Health (Grand Rapids, MI) \cite{Murugesan2023} that were assessed by a radiologist to be reasonable. We computed the same overlap and distance metrics between our baseline method segmentations and the ones available in IDC \cite{Fedorov2023}. For the second comparison, we used a pre-trained model available from the nnU-Net \cite{Isensee2021}. The nnU-Net framework provides a large number of pre-trained models, including a model trained on CHAOS MR data for segmentation of the liver, spleen, left and right kidneys. We compare the performance of our method to the nnU-Net model on the AMOS MR dataset, using expert annotated ground-truth segmentations available for AMOS MR.  \\
\indent Table ~\ref{Table4} displays the quantitative results between our baseline approach and the BAMF method on IDC data. AMOS MR results are also shown for the evaluated nnU-Net model and our baseline method, compared to expert annotations. We observe a high overlap between the liver segmentations from BAMF and our baseline segmentations, on IDC MR TCGA-LIHC data (0.92 ± 0.02 DSC). On AMOS MR, we can observe comparable results to the nnU-Net pre-trained model, when evaluated against expert annotations (0.91 ± 0.08 vs 0.91 ± 0.03 DSC for the left kidney, ours and nnU-Net, respectively). Please refer to the supplementary material for qualitative results of the evaluated methods.

\begin{table}[t]
\caption{Quantitative results comparing publicly available methods to our approach using the Dice score and HD95. 
% We compare the output of the BAMF annotations \cite{Murugesan2023} to our predictions. We also compare the performance of a pre trained nnU-Net model \cite{Isensee2021} to our predictions on the AMOS MR dataset, using expert annotations as ground-truth for metrics computation. 
% Dice score and average Hausdorff distance (95th percentile), along with the standard deviation values are shown.
}\label{Table4}
\centering
\begin{tabular}{|c|c|c|c|c|}
\hline
Method & Dataset & Segment & Dice & HD95 \\
\hline
BAMF \cite{Murugesan2023} & TCGA-LIHC (MR) & Liver & 0.92 ± 0.02 & 7.30 ± 2.67 \\
\hline
nnU-Net model \cite{Isensee2021} & AMOS MR & Liver & 0.83 ± 0.23 & 31.04 ±  57.29 \\
nnU-Net model \cite{Isensee2021} & AMOS MR & Spleen &  0.88 ± 0.20 & 14.99 ± 52.49 \\
nnU-Net model \cite{Isensee2021} & AMOS MR & Right kidney & 0.87 ± 0.20 & 16.04 ± 40.12 \\
nnU-Net model \cite{Isensee2021} & AMOS MR & Left kidney & 0.91 ± 0.03 & 7.62 ± 1.82 \\
\hline
\end{tabular}
\end{table}
%%%%%%%%%%%%%%%%%%%%%%%%%%%%%%%%%%%%%%%%%%%%%%%%%%%%%%%%%%%%%%%%%%%%%%%%
\section{Conclusion}
We proposed a modality-agnostic deep learning method for abdominal organ segmentation using a domain randomization strategy trained on CT label maps. Our method shows promising results when validating and testing on publicly available datasets, as well as a comparison to publicly available fully-supervised segmentation methods. Additionally, we performed an ablation study to understand the effect of prediction of multiple labels and the addition of clustering. Our baseline method, including additional labels and generation-only foreground and background clustering labels, performed the best for both the overlap and distance metrics, namely Dice score and Hausdorff distance (95th percentile), compared to the ablation studies methods. Our study presents a few limitations, as abdominal structures outside of the liver, spleen and kidneys did not perform well. This could be due to the higher heterogeneity in those regions with regards to texture, appearance and location, and requires further investigation. Future work includes modification of the training data to better represent MR-specific features, and refinement of the ground truth labels in order to offer a better label anatomical consensus between collections. 

\bibliographystyle{splncs04}
\bibliography{references.bib}
%





% LNCS Homepage, \url{http://www.springer.com/lncs}, last accessed 2023/10/25
% \end{thebibliography}

\end{document}

\begin{figure}[h]
\includegraphics[width=1\textwidth, trim={0cm 8cm 0cm 5.5cm},clip]{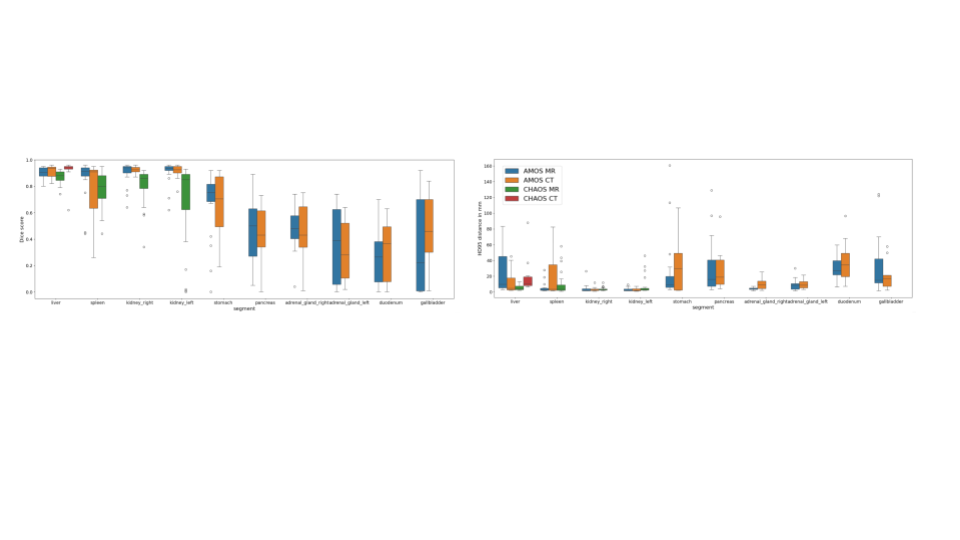}
\centering
\caption{Boxplots for the distribution of Dice scores (left) and the distribution of Hausdorff distance values (95th percentile) for all segments, between our proposed method and the expert annotations for the AMOS and CHAOS collections. Note the high agreement for the liver, spleen and kidneys, and the low agreement for the others. This may be due to the reduced ability of the network to capture variable appearance of those regions. Predicted segments with undefined Hausdorff distance values (due to empty segmentation sets) were removed from all of the metrics calculation. 9 segment predictions for AMOS MR/CT.} \label{FigureS1}
\end{figure}

\begin{table}[H]
\caption{Table for the distribution of Dice scores (left) and the distribution of Hausdorff distance values (95th percentile) for all segments, for the AMOS and CHAOS  collections. Subjects with undefined Hausdorff distance values were removed for the overall metrics calculation. Predicted segments with undefined Hausdorff distance values (due to empty segmentation sets) were removed from all of the metrics calculation. 9 segment predictions for AMOS MR and 9 segments for AMOS CT were removed. }\label{TableS1}
\centering
\begin{tabular}{|c|c|c|c|c|c|}
\hline
Dataset & Segment & MR & MR & CT & CT \\
\hline
& & Dice & HD95 & Dice & HD95 \\
\hline
AMOS & Liver & 0.90 ± 0.04 & 25.13 ± 26.05 & 0.91 ± 0.05 & 13.15 ± 15.43 \\ %&  
AMOS & Spleen & 0.86 ± 0.15 & 5.34 ± 6.37 & 0.78 ± 0.22 & 19.37 ± 26.20 \\ %& 
AMOS & Right kidney & 0.90 ± 0.08 & 4.24 ± 5.50 & 0.91 ± 0.05 & 2.83 ± 1.84 \\ %&
AMOS & Left kidney & 0.91 ± 0.08 & 2.92 ± 2.04 & 0.92 ± 0.03 & 3.55 ± 2.84 \\ %&
AMOS & Stomach &  0.68 ± 0.24 & 25.38 ± 39.55 & 0.64 ± 0.24 & 34.06 ± 30.26 \\ %&
AMOS & Pancreas &  0.46 ± 0.23 & 30.62 ± 34.36 & 0.44 ± 0.21 & 27.97 ± 25.82 \\ %&
AMOS & Duodenum &  0.27 ± 0.21 & 31.32 ± 15.53 & 0.30 ± 0.22 & 37.88 ± 25.63 \\ %&
AMOS & Gallbladder &  0.34 ± 0.33 & 33.66 ±38.24  & 0.46 ± 0.27 & 21.19 ± 18.69 \\ %&
AMOS & Right adrenal gland & 0.47 ± 0.15 & 4.21 ± 1.42 & 0.45 ± 0.22 & 9.89 ± 7.04 \\ %&
AMOS & Left adrenal gland &  0.36 ± 0.27 & 9.19 ± 7.09 & 0.31 ± 0.23 & 9.99 ± 6.22 \\ %&
\hline
CHAOS & Liver & 0.87 ± 0.05 & 5.77 ± 3.27 & 0.91 ± 0.10 & 21.08 ± 23.95 \\%&
CHAOS & Spleen  & 0.78 ± 0.13 & 10.46 ± 15.06 & --- & --- \\ %&
CHAOS & Right kidney & 0.80 ± 0.14  & 3.53 ± 2.29 & --- & --- \\ %&
CHAOS & Left kidney & 0.68 ± 0.31 & 7.79 ± 11.19 & --- & --- \\ %&
\hline
\end{tabular}
\end{table}

\begin{figure}[h]
\includegraphics[width=0.8\textwidth, trim={0cm 6cm 0cm 6cm},clip]{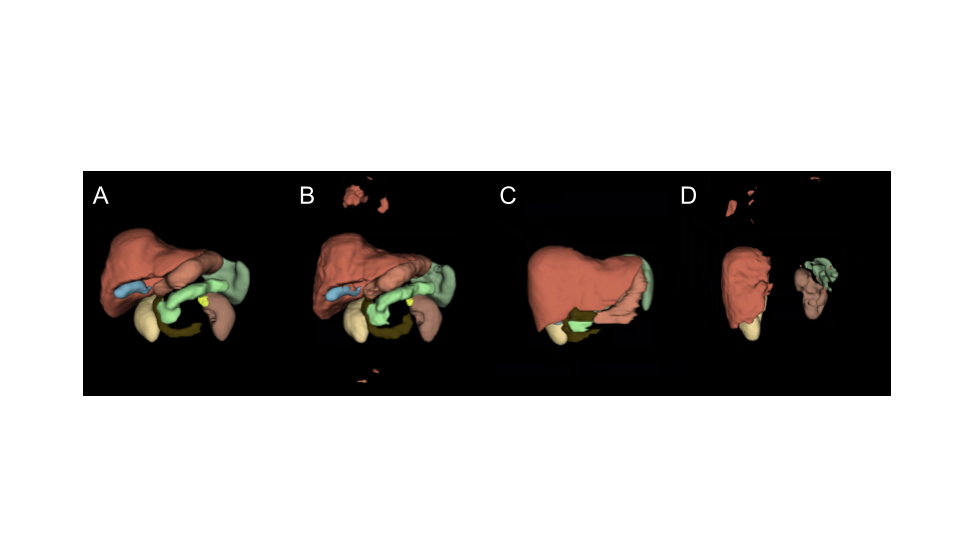}
\centering
\caption{Comparison of the segmentation results of the expert annotations to the proposed method, for two sample subjects from AMOS MR. A: Expert annotations; B: Proposed method on the subject as (A) with high agreement; C: Expert annotations; D: Proposed method on the same subject as (C) with low agreement.} \label{FigureS2}
\end{figure}

\begin{figure}[h]
\includegraphics[width=0.7\textwidth, trim={0cm 1.5cm 0cm 0cm},clip]{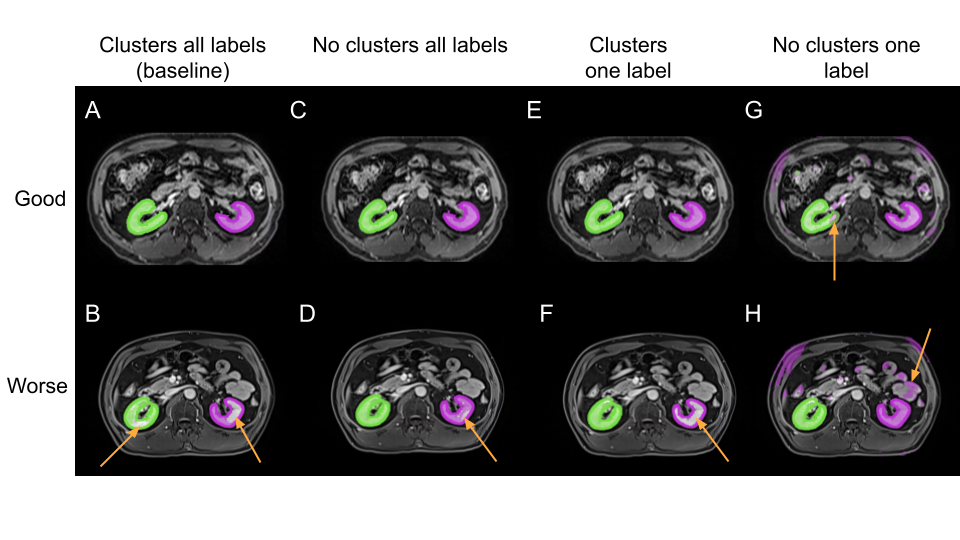}
\centering
\caption{displays the qualitative results of the kidney segmentation for the baseline method along with the three ablation studies models. The ground truth expert segmentations are given by the thick segment boundary line, while the AI predictions are filled. The right kidney is green, and the left kidney is purple. Yellow arrows indicate missing segments or incorrect segmentations. Note the higher overlap of the expert segmentations versus our AI models for methods that include all labels. } \label{FigureS3}
\end{figure}

\begin{figure}[h]
\includegraphics[width=0.7\textwidth, trim={0cm 0.5cm 0cm 4cm},clip]{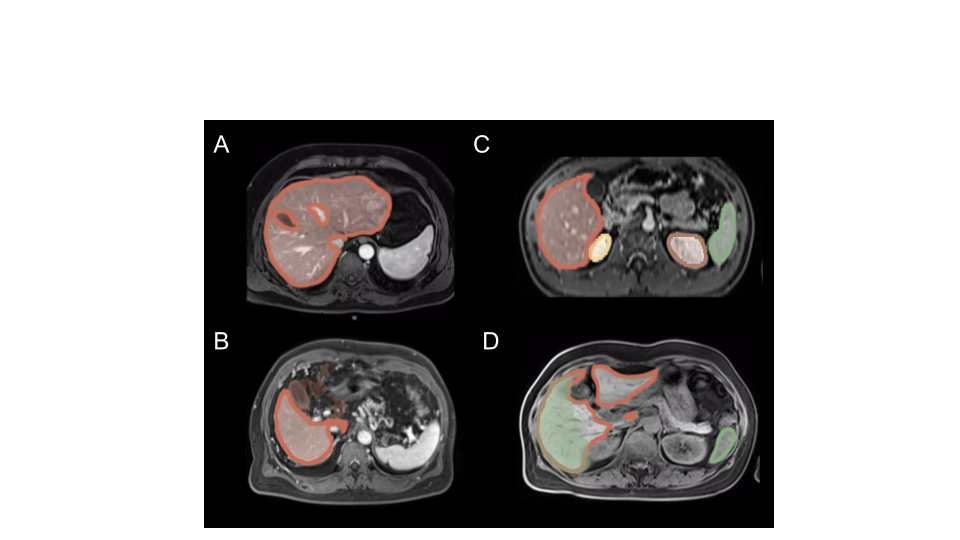}
\centering
\caption{Comparison of the proposed method with publicly available methods. A) Proposed method (filled) vs results from the BAMF method for liver segmentation (border) for a subject with high agreement. B) Proposed method (filled) vs results from the BAMF method for liver segmentation (border) for a subject with low agreement. C) Results from the nnU-Net model (filled) vs the ground truth (border) for a subject with high agreement. D) Results from the nnU-Net model (filled) vs the ground truth (border) for a subject with low agreement. Note the model predicts the liver (red) as the spleen (green) and vice versa.} \label{FigureS4}
\end{figure}